\documentclass{article}
\usepackage{graphicx} %
\usepackage{comment}
\usepackage{geometry}
\usepackage{color}
\usepackage{float}
\usepackage{caption, subcaption}
\usepackage{amsmath}
\usepackage{amsfonts}
\usepackage{mathrsfs}  %
\usepackage{amssymb}
\usepackage{graphicx}

\usepackage[backend=biber, style=apa, natbib=true, uniquelist=false, uniquename=false, sorting=nyt, sortcites=true]{biblatex}
\letbibmacro{orig-doi+url}{doi+url}
\renewbibmacro*{doi+url}{%
	\iffieldundef{doi}%
	{\iffieldundef{url}%
		{\printfield{howpublished}}%
		{\usebibmacro{orig-doi+url}}%
	}%
	{\usebibmacro{orig-doi+url}}%
}%

\AtEveryBibitem{\clearfield{month}}
\AtEveryCitekey{\clearfield{month}}
\AtEveryBibitem{\clearfield{labelmonth}}
\AtEveryCitekey{\clearfield{labelmonth}}
\AtEveryBibitem{\clearfield{day}}
\AtEveryCitekey{\clearfield{day}}
\AtEveryBibitem{\clearfield{labelday}}
\AtEveryCitekey{\clearfield{labelday}}

\DeclareBibliographyDriver{customa}{%
  \printfield{usera}%
  \finentry
}

\usepackage{appendix}
\usepackage{tablefootnote}  %
\usepackage[normalem]{ulem}   %
\usepackage{bm}
\usepackage{bbm}
\usepackage{enumitem}
\usepackage{setspace}
\usepackage{booktabs} %
\usepackage{todonotes}
\usepackage{mathtools}
\usepackage{multirow}
\usepackage[amsmath,thmmarks,standard]{ntheorem}

\let\originalleft\left
\let\originalright\right
\renewcommand{\left}{\mathopen{}\mathclose\bgroup\originalleft}
\renewcommand{\right}{\aftergroup\egroup\originalright}

\DeclarePairedDelimiter\abs{\lvert}{\rvert}%
\DeclarePairedDelimiter\norm{\lVert}{\rVert}%
\makeatletter
\let\oldabs\abs
\def\abs{\@ifstar{\oldabs}{\oldabs*}}
\let\oldnorm\norm
\def\norm{\@ifstar{\oldnorm}{\oldnorm*}}
\makeatother

\makeatletter
\renewtheoremstyle{plain}%
  {\item[\hskip\labelsep \theorem@headerfont ##1\ ##2\theorem@separator]}%
  {\item[\hskip\labelsep \theorem@headerfont ##1\ ##2 \mdseries (##3)\theorem@headerfont \theorem@separator]}
\makeatother

\theoremseparator{.}
\renewtheorem{theorem}{Theorem}[section]

\renewtheorem{lemma}[theorem]{Lemma}

\theoremclass{Theorem}
\theoremstyle{plain}
\theorembodyfont{\normalfont}
\theoremheaderfont{\bf}
\theoremseparator{.}
\theoremsymbol{\enskip\ensuremath{\diamond}}
\renewtheorem{definition}[theorem]{Definition}
\renewtheorem{remark}[theorem]{Remark}
\renewtheorem{example}[theorem]{Example}

\theoremstyle{nonumberplain}
\theoremheaderfont{\itshape}
\theorembodyfont{\normalfont}
\theoremseparator{.}
\theoremsymbol{\ensuremath{\square}}
\renewtheorem{proof}{Proof}

\makeatletter
\renewtheoremstyle{nonumberplain}%
  {\item[\hskip\labelsep \theorem@headerfont ##1\theorem@separator]}%
  {\item[\hskip\labelsep \theorem@headerfont ##3\theorem@separator]}
\makeatother

\newif\ifEndMarkerWrong

\AddToHook{env/proof/end}{%
  \ifEndMarkerWrong
    \ifvmode
        \PackageWarning{ntheorem}{End of proof in vertical mode\MessageBreak
        Check QED marker}%
    \fi
  \fi}
  
\AddToHook{env/example/end}{%
  \ifEndMarkerWrong
    \ifvmode
        \PackageWarning{ntheorem}{End of proof in vertical mode\MessageBreak
        Check QED marker}%
    \fi
  \fi}
  
\AddToHook{env/definition/end}{%
  \ifEndMarkerWrong
    \ifvmode
        \PackageWarning{ntheorem}{End of proof in vertical mode\MessageBreak
        Check QED marker}%
    \fi
  \fi}

\numberwithin{equation}{section}

\newcommand{\db}{\mathfrak{d}}
\newcommand{\dbp}{\mathfrak{p}}
\newcommand{\dbf}{\mathfrak{f}}
\newcommand{\dbs}{\mathfrak{s}}
\newcommand{\dbr}{\mathfrak{r}}

\usepackage{parskip}
\usepackage[unicode=true,bookmarks=true,bookmarksnumbered=true,bookmarksopen=true,breaklinks=true,pdfborder={0 0 0},pdfborderstyle={},colorlinks=true,allcolors=blue,hypertexnames=false]{hyperref}
\usepackage{bookmark}

\makeatletter
\let\oldsection\section
\renewcommand{\section}{\@ifstar{\sectionstar}{\oldsection}}
\NewDocumentCommand{\sectionstar}{s o m}{%
	\oldsection*{#3}%
	\addcontentsline{toc}{section}{#3}%
}

\let\oldsubsection\subsection
\renewcommand{\subsection}{\@ifstar{\subsectionstar}{\oldsubsection}}
\NewDocumentCommand{\subsectionstar}{s o m}{%
	\oldsubsection*{#3}%
	\addcontentsline{toc}{subsection}{#3} %
}
\makeatother

\renewcommand{\epsilon}{\varepsilon}

\title{The Complexities of Differential Privacy for Survey Data\footnote{Forthcoming in \fullcite{hotzDataPrivacyProtection2026}}} %

\author{J\"org Drechsler\footnote{joerg.drechsler@iab.de \\ \indent\ \ Institute for Employment Research, Nuremberg, Germany; Ludwig-Maximilians-Universit\"at, Munich, Germany; University of Maryland, College Park, USA.}\ \ \& James Bailie\footnote{bailie@chalmers.se \\ \indent\ \ Department of Statistics, Harvard University, Cambridge, USA.}}

\date{}%

\begin{document}

\maketitle

\begin{abstract}
The concept of differential privacy (DP) has gained substantial attention in recent years, most notably since the U.S. Census Bureau announced the adoption of the concept for its 2020 Decennial Census. However, despite its attractive theoretical properties, implementing DP in practice remains challenging, especially when it comes to survey data.
In this chapter we present some results from an ongoing project funded by the U.S. Census Bureau that is exploring the possibilities and limitations of DP for survey data. %
Specifically, we identify five aspects that need to be considered when adopting DP in the survey context: the multi-staged nature of data production; 
the limited privacy amplification from complex sampling designs; the implications of survey-weighted estimates; the weighting adjustments for nonresponse and other data deficiencies, and the imputation of missing values. 
We summarize the project's key findings with respect to each of these aspects and also discuss some of the challenges that still need to be addressed before DP could become the new data protection %
standard at statistical agencies. %
\end{abstract}

\section{Introduction}

Differential privacy (DP) \citep{dwork2006calibrating} has become the quasi--gold standard in recent years for data collection and dissemination whenever privacy or confidentiality is a concern. It offers formal (that is, mathematically quantifiable) privacy guarantees by bounding the influence that any single record of the database can have on the computed outputs. The fundamental difference to earlier privacy frameworks such as $k$-anonymity is that the guarantees are a property of the mechanism generating the output and not a property of the data. DP specifies how much noise the mechanism needs to introduce to ensure that the probability of obtaining a specific output does not change substantially if one record in the database is changed. In simple examples where we are interested in creating a DP version of an unprotected statistic such as the sample mean, the required amount of noise depends on two components. The first component is the privacy loss parameter $\varepsilon$ (also called the privacy loss `budget'), which determines how much the probability of obtaining a specific result is allowed to change.\footnote{For simplicity we limit our exposition to the classical bounded $\varepsilon$-DP setting, where `bounded' means that neighboring datasets are defined as datasets that can be obtained by replacing a single record with another record without changing the size of the database. Similar arguments would apply for other variants, such as $(\varepsilon,\delta$)-DP, $\rho$-zero-concentrated DP, or $f$-DP, and for other definitions of neighboring datasets, such as unbounded DP for which a neighboring database is obtained by adding or removing a single record.} The smaller the privacy loss parameter, the more noise needs to be added and the better the level of protection offered. The second component is the sensitivity of the unprotected statistic of interest, which is the maximum possible change of the statistic when changing one record in the database. The higher the sensitivity, the more noise needs to be added. %

To illustrate, we can look at one of the classical DP mechanisms that is often used as a building block in more complex algorithms: the Laplace mechanism, which, for any univariate statistic $f$, ensures $\varepsilon$-DP by adding a random draw from a Laplace distribution centered at zero with scale parameter $b=\Delta f/\varepsilon$. The parameter $\Delta f$ is the sensitivity of $f$ measured as the maximum absolute distance (the $L_1$ norm) of the statistic computed over two neighboring datasets, i.e., two datasets that differ only in a single record. 
With this mechanism, the dependence on the two parameters is obvious: More noise is added for outputs with higher sensitivity and smaller values of $\varepsilon$.

This is one of the attractive properties of DP. The concept is very intuitive and requires only three steps, which in principle seem straightforward to apply: (i) define the maximum privacy loss that is still considered acceptable and select a value for $\varepsilon$ accordingly; (ii) identify the sensitivity of the statistic of interest (for example, the sensitivity of a proportion under bounded $\varepsilon$-DP is simply $1/n$, where $n$ is the number of records in the database); and (iii) choose a DP mechanism that infuses the right amount of noise into the reported output based on the parameters from steps (i) and (ii). %
Of course, in practice all three steps have their challenges. The discussion on how to choose and interpret the privacy loss parameter shows no signs of abating \citep{hsuDifferentialPrivacyEconomic2014, dworkDifferentialPrivacyPractice2019, abowdEconomicAnalysisPrivacy2019, tschantz2020sok, nanayakkara2023chances, drechslerDifferentialPrivacyGovernment2023,bailieRefreshmentStirredNot2024a}%
; the sensitivity of the output is not always easy to compute and can be unbounded without further assumptions \citep{casacubertaWidespreadUnderestimationSensitivity2022}; and finding a suitable DP mechanism can be challenging. Besides, there are often some hidden complications to DP in practice beyond what this three-step process makes apparent \citep{abowd2022topdown, seemanPrivacyUtilityDifferential2023, cummingsAdvancingDifferentialPrivacy2024}. (For example, for the same research question there can be multiple choices for which statistic is used in step (ii), and it can be difficult to determine which one leads to the most efficient DP mechanism.) %
Still, the three components remain the same across applications and at least the general setup is well defined.

However, when working with survey data, there are additional complexities that typically do not arise in other settings. Moreover, the implications of using DP in the context of surveys have received little attention in the DP literature until recently. This led the U.S. Census Bureau to conclude in 2022 that ``the science does not yet exist'' to implement DP in its American Community Survey \citep{uscensusbureauDisclosureAvoidanceProtections2022}. An expert panel convened by the National Academies of Sciences, Engineering, and Medicine reached a similar conclusion with respect to the Survey of Income and Program Participation \citep{raghunathanRoadmapDisclosureAvoidance2023}.

Given its commitment to formal privacy for all its data products, including its surveys \citep{uscensusbureauProtectingConfidentialityAmerica2018}, the U.S. Census Bureau sponsored a research project from 2020 to 2025 to better understand the complexities that arise when adopting DP in the survey context. In this chapter, we will summarize some of the key findings of this project and also discuss some of the challenges that still need to be addressed. Overall, we identify five aspects that need to be considered when implementing DP in the survey context:
\begin{itemize}
    \item Data production is a multistage process. As such, there are various options for how and where to integrate DP in this pipeline, each of which come with their own advantages and disadvantages.%
\item Previous studies found that sampling can amplify DP's privacy guarantees. However, these amplification effects do not necessarily hold for the complex sampling designs used by statistical agencies.
\item %
These complex sampling designs need to be incorporated into any survey statistic and hence must also be incorporated into any DP mechanism.
\item Weighting adjustments are routinely used to account for unit nonresponse and to benchmark to known population totals. %
As these adjustments can substantially increase the sensitivity of the survey statistic, there is a need to develop robust adjustment strategies that are congenial to DP. 
\item Item nonresponse is often addressed using imputation, but similar to weighting adjustments some standard imputation techniques can greatly inflate the sensitivity of the resulting statistic. Ongoing research is currently investigating the feasibility of differentially private imputation techniques.
\end{itemize}
We will discuss each of these aspects in the remainder of this chapter.

\section{DP and the Multistage Process of Data Production}

\subsection{The Survey Pipeline}

\begin{figure}
    \centering
    \includegraphics[width=\linewidth]{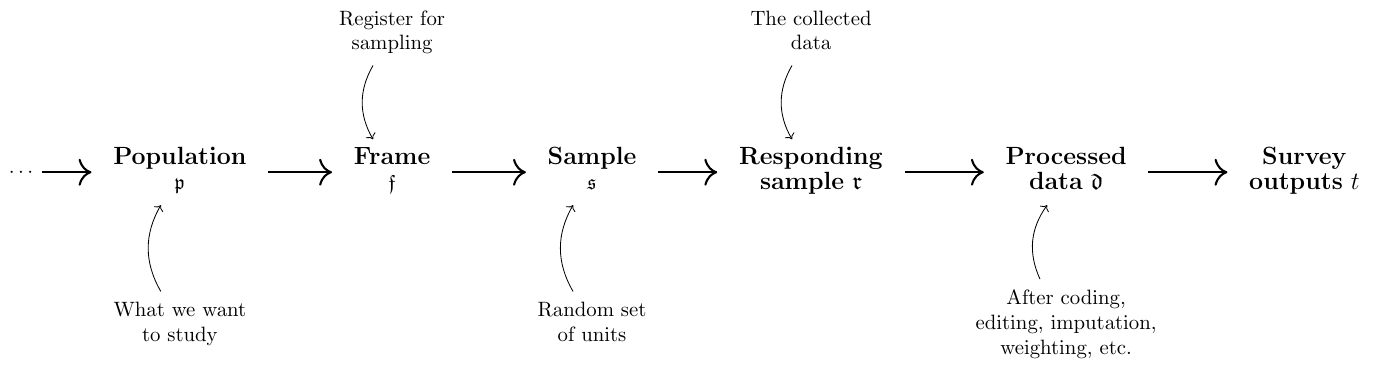}
    \caption{A \emph{survey pipeline} consists of multiple steps, of which some of the most important are: determining the target population to be studied; constructing the frame; drawing the sample; collecting survey data from the responding units; processing the data (including coding free-form responses; editing inconsistent or improbable data; imputing missing records or variables; calculating the survey weights; and injecting privacy-protecting noise); and computing the survey outputs. There are of course additional steps to a survey pipeline after the survey outputs are released (such as data analysis) but, as they are not important to this chapter's subject, we exclude these steps from discussion. While not shown in this figure, it should be noted that data from previous stages of a pipeline are often used in later stages. (For example, the frame is usually used in computing the survey weights during the production of the processed data.)}
    \label{figPipeline}
\end{figure}

The production of survey data is a complex multistage process (Figure~\ref{figPipeline}). The design of a survey typically begins by conceiving the \emph{target population}: the set of units that one wants to study. Usually, the target population is not actually specified as a concrete list of units. Instead, it is defined conceptually: ``all adults in Massachusetts'' or ``all businesses in Hawaii.'' Once the target population has been defined, the \emph{frame} is sourced. The frame is a register of units from which the sample will eventually be drawn. It must include sufficient contact information so that the sampled units can be surveyed. The frame should align with the target population as much as possible. However, perfect alignment is not possible in most cases, even when the target population and the frame have the same inclusion criteria, because errors will typically be made in the frame's construction. These errors will result in overcoverage (including units that are not in the target population) and undercoverage (not including units that are in the target population).

A \emph{sample} is randomly drawn from the frame according to the survey's \emph{sampling design}: %
the probability distribution that specifies for every potential sample the chance that sample is selected. %
After sample selection, the statistical agency will solicit survey data from the sampled units. Most surveys, especially modern ones, suffer from nonresponse. This means only a subset of the sampled units will respond and the agency will not obtain survey data from the other units. Data collected from the responding sample, along with the frame and some auxiliary information (such as data from administrative records or from previous censuses or surveys), are passed through a number of complex data processing steps before the survey outputs are computed and released. These data processing steps often include editing survey responses to correct errors in data recording; coding each free-form answer into a categorical variable; imputing missing answers to individual survey questions (``item nonresponse'') or to the entire survey questionnaire (``unit nonresponse''); and calculating multiple sets of survey weights for each record---to account for unequal probabilities of selection in the sampling design, to mitigate bias due to nonresponse patterns, and to calibrate survey data to auxiliary sources of information. %
Finally, we note that data may be deliberately injected with artificial noise at any point in the survey pipeline, so that releasing the survey outputs does not breach the privacy of the data subjects. %

\subsection{DP in the Survey Pipeline}\label{DPInPipeline}

DP is a criterion applied to \emph{data release mechanisms}: algorithms %
that take data as input and produce %
a set of outputs that will then be published (that is, ``released''). %
(For a more extensive, yet still approachable, discussion on what DP is, we direct the reader to the companion chapter \cite{bailieRefreshmentStirredNot2025a} in this volume.)
Implementing DP involves both %
designing a data release mechanism that is compliant with DP, as well as integrating that mechanism into the relevant data pipeline. %
Both tasks are crucial for successfully producing outputs with high accuracy and good privacy protection.

\begin{figure}
    \centering
    \begin{subfigure}[b]{0.49\textwidth}
        \centering
        \includegraphics[width=\textwidth]{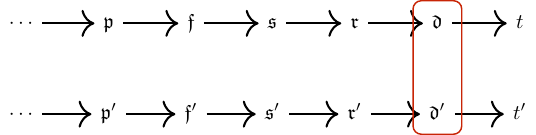}
        \caption{}
        \label{figExamplePipeline1}
    \end{subfigure}
    \hfill
    \begin{subfigure}[b]{0.49\textwidth}
        \centering
        \includegraphics[width=\textwidth]{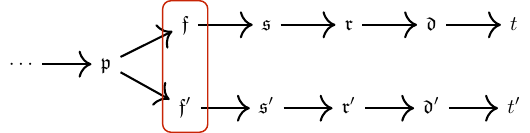}
        \caption{}
        \label{figExamplePipeline2}
    \end{subfigure}
    \begin{subfigure}[b]{0.49\textwidth}
        \centering
        \includegraphics[width=\textwidth]{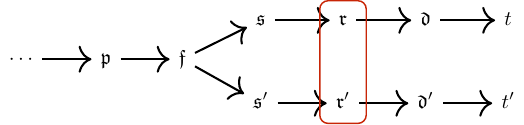}
        \caption{}
        \label{figExamplePipeline3}
    \end{subfigure}
    \caption{Three examples of where to start the data release mechanism (circled in red) in the survey pipeline and which of the previous stages to take as invariant (those stages before the pipeline branches). Recall from Figure~\ref{figPipeline} that $\dbp$ denotes the population, $\dbf$ the frame, $\dbs$ the sample, $\dbr$ the responding sample, $\db$ the processed data and $t$ the survey outputs. The apostrophe $'$ indicates an alternative realisation of the associated variable. Figure~(a) illustrates the standard approach in which there are no invariants and the data release mechanism only executes the final step of the survey pipeline---transforming the processed data into the survey outputs. In Figure (b), the mechanism begins with the frame and includes the sampling, responding and processing steps. The population is considered invariant. In Figure (c), the mechanism takes as input the responding sample. Both the population and the frame are taken as invariant, so that DP only compares samples from the same frame. This reduces the sensitivity of weighted estimators at the expense of less privacy (Section~\ref{sectionSurveyWeights}).}
    \label{figExamplesPipeline}
\end{figure}

There are two important considerations when integrating a DP mechanism into a data pipeline. Firstly, at what point in the pipeline should the DP mechanism \emph{start}? And secondly, which of the earlier stages of the data pipeline should be considered \emph{invariant}---i.e., should be treated as fixed---by DP? With survey pipelines, there are a number of possible options with respect to both considerations. %
In the option most commonly seen in the DP literature, the data release mechanism starts at the end of the pipeline and %
performs just the last step---computing the survey outputs from the processed data---and none of the previous steps are taken as invariant (Figure~\ref{figExamplePipeline1}). However, a mechanism could conceivably start at any point of the survey pipeline and incorporate all the steps that follow. For example, it could take as input the frame, execute the sampling step, process the data and finally compute the survey outputs (Figure~\ref{figExamplePipeline2}). Furthermore, any of the steps before the mechanism starts could conceivably be taken as invariant. (For explanations of how steps of the data pipeline can be taken as invariant by DP, or how a mechanism's starting point can be encoded into DP, we again direct the reader to \cite{bailieRefreshmentStirredNot2025a}.) In the rest of this section, we will explore these two considerations %
in turn.

Throughout this chapter, we assume that the data release mechanism always includes the final step of the survey pipeline, the computation of the survey outputs.%
\footnote{Technically, a data release mechanism is simply %
an algorithm that takes data as input and outputs some (possibly noisy) transformation of that data. So, in principle, a data release mechanism could be incorporated into a survey pipeline even if it ends before the final step of the pipeline. 
(And such a mechanism could still be compliant with DP.) In this case, the survey pipeline includes additional post-processing steps after %
the data release mechanism ends but before the computation of the outputs that will be published. Such post-processing steps are usually included to improve the utility, usability or accessibility of the survey outputs. On the other hand, any data release mechanism can always be extended to one that ends with the final step of the survey pipeline, and any DP guarantees afforded to the original mechanism automatically carry over to the extended one by the post-processing theorem. (The post-processing theorem states that any function of a DP mechanism's output---i.e., any ``post-processing''---also satisfies DP with at most the same privacy loss.) 
Therefore, we do not gain anything by considering DP mechanisms that end before the survey pipeline's final step.
} Under this assumption, a survey pipeline can be split into those steps that are executed before the data release mechanism starts and those steps that are executed by the mechanism. Yet choosing where to make this split is not a simple matter. In fact, there are a number of complexities associated with starting the data release mechanism earlier or later in the pipeline. We identify five. 

Firstly, starting the DP mechanism earlier can complicate the computation of the cumulative privacy loss across multiple data release mechanisms because DP's composition theorems\footnote{A composition theorem describes how to bound the total privacy loss incurred by multiple DP data releases that are all based on the same confidential dataset. For example, the composition theorem for pure $\varepsilon$-DP states that: if there are $K$ mechanisms $M_1,\ldots, M_K$ that all satisfy pure $\varepsilon$-DP and all have the same input dataset, then the total privacy loss---that is, the privacy loss of the mechanism that publishes all the outputs of $M_1, \ldots, M_K$ together---is bounded by the sum $\sum_k \varepsilon_k$ over the privacy losses $\varepsilon_k$ of each mechanism $M_k$. Existing composition theorems assume that the noise added by each mechanism is ``fresh'', i.e., independent of everything else.} %
are not applicable when there is dependence between the mechanisms' noise terms (which can happen, for example, when their sampling designs are dependent) \citep{drechslerWhoseDataIt2024}.

Secondly, as we will describe in Section~\ref{sectionSampling}, including the sampling step within the data release mechanism can amplify DP's privacy guarantees without degrading data utility. However, this privacy amplification can be nullified if the attacker knows that the record they are attacking is in the sample \citep{drechslerWhoseDataIt2024}. More generally, if the attacker has knowledge about information intermediary to the DP mechanism (that is, information that is conditionally dependent on confidential data, or on the artificial noise introduced by the mechanism, conditioning on the output of the mechanism), the privacy guarantees afforded by DP can be weakened. For this reason, DP prohibits the direct release of such information. Therefore, because the choice of the sampling design is often dependent on data in the frame, %
the sampling design cannot be directly made public but instead can only be released by including it in the set of DP-protected survey outputs. %

However, defining a data release mechanism---let alone one that satisfies DP---that releases the sampling design is challenging due to the third complexity we identify: Incorporating existing steps of a survey pipeline into a data release mechanism can be difficult. %
A data release mechanism is an algorithm that must be fully specified in order to be analysed by DP; hence any stage of the survey pipeline must first be fully ``algorithmized'' (that is, the process by which each of the stage's possible inputs is transformed into one of its outputs must be completely %
and programmatically specified) before it can be included in a mechanism.\footnote{The post-processing theorem provides an exception to this general rule. If the preliminary steps of a data release mechanism, taken on their own, satisfy DP, then the later steps of the mechanism do not need to be algorithmized, because the post-processing theorem ensures that the mechanism as a whole always satisfies DP regardless of what the later steps do. All that must be checked is that the later steps only use the DP outputs from the preliminary steps, and not some other data. However, this exception does not apply to the survey pipeline steps under discussion (choosing a sampling design, coding and editing) because these steps are typically applied before---not after---privacy protection.} A survey pipeline often includes a number of complex, ill-defined and human-intensive tasks, such as %
building the frame, choosing a sampling design, coding and editing. Because these tasks all usually require a degree of human judgment, they would be difficult to algorithmize. %
Moreover, including these procedures---or other procedures often found in a survey pipeline---in a data release mechanism can add difficulties to making the mechanism compliant with DP. (In later sections, we will discuss some such difficulties as they relate to the weighting and imputation procedures.%
)

Fourthly, even if a data release mechanism begins later in the survey pipeline so that some steps of the pipeline do not have to be incorporated in the mechanism, implementing DP still requires understanding those steps' effect on the mechanism's input data. For example, some imputation techniques replace missing records with copies of nonmissing donor records. This means an individual survey respondent can contribute to multiple records in the post-imputation dataset. This complicates the appropriate definition of neighboring datasets, since there is no longer an exact correspondence between the dataset's records and the real-world entities (the individual respondents) that should be protected: In the post-imputation dataset, changing a single record does not correspond to changing the data of one entity. Hence, na\"ively applying DP to the post-imputation dataset will not provide a donor record with the expected level of protection; that is, the privacy guarantees for a donor record will be weaker than those for a post-imputed record. In general, the later the DP mechanism begins, the more difficult it is to determine an appropriate notion of neighbors since steps earlier in the pipeline may introduce dependencies between dataset records, thereby complicating the relationship between records and data subjects.%

Fifthly, and most fundamentally, the starting point of the data release mechanism determines what form of the data is protected by that mechanism \citep{bailieRefreshmentStirredNot2024a}. For example, if a DP mechanism begins after data processing, then it is the processed data---and not, for example, the raw responses from the data providers---that are protected by that mechanism. That is to say, DP guarantees implicitly assume that the attacker is interested in inferring the data that is input into a DP mechanism. Measures of protection are in terms of the attacker's ability to learn this input data---and not the data at other points in the pipeline. 
If the DP mechanism takes the processed data as input, then the DP guarantees apply to the processed data and %
do not necessarily carry over to the responding sample data. %
In order to have guarantees for the responding sample, the statistical agency must show that the pipeline from the responding sample to the survey outputs (considered as a data release mechanism) also satisfies DP.

These five complexities demonstrate that there can be conflicting demands in deciding where a DP mechanism should start within the survey pipeline. For example, suppose a statistical agency wants to protect the unprocessed survey responses. Then either the coding and editing steps will need to be included in the agency's mechanism (which may be difficult because these steps could be hard to algorithmize)  or these steps will need to be removed from the survey pipeline (which could decrease the quality of the survey outputs).

We now return to the question of which steps of the survey pipeline should DP take as invariant. DP assesses the privacy of a data release mechanism by comparing the survey outputs' distribution under pairs of counterfactual input datasets. These input datasets are generated by counterfactual runs of the initial steps of the pipeline, up until the data release mechanism begins. By taking some of these steps as invariant, DP's counterfactual comparisons are reduced to only those pairs of input datasets that share the same realization of the invariant steps. For example,  suppose the steps in the survey pipeline that generate the population and the frame are taken as invariant and the data release mechanism starts with the responding sample (Figure~\ref{figExamplePipeline3}). %
Then DP only compares those responding samples (i.e., those counterfactual input datasets) that could have come from the same frame. Adding invariants will weaken the privacy guarantees provided by DP \citep{kiferBayesianFrequentistSemantics2022, abowd2022topdown, bailieRefreshmentStirredNot2024b}. In general, the later the stage of the pipeline that is kept invariant, the greater the reduction in privacy. However, invariants may be justifiable when the output of the invariant steps can be considered as public knowledge (such as if the frame was sourced commercially rather than constructed from confidential information). Moreover, constraining some steps to be invariant has the advantage of reducing the sensitivity of weighted estimators and thereby decreasing the noise that must be added for privacy protection (Section~\ref{sectionSurveyWeights}). %

\section{DP with Complex Sampling Designs}\label{sectionSampling}

Statistical agencies have been aware for decades that sampling can be a simple and effective strategy to reduce disclosure risks simply because an attacker can no longer be sure whether a specific target record is included in the sample or not. This is the main motivation why most statistical agencies only release samples from their censuses as public use micro datasets (they typically also apply additional measures to further increase the level of protection). This idea has been formalized in several papers in the context of DP \citep{KasiviswanathanLNRS11,WangLF16b,BunNSV15, balle2018privacy,WangBK19}. The authors show that the level of privacy is amplified through sampling, i.e., the actual privacy guarantees are higher then those implied by the chosen privacy loss parameters when protecting the sample output. %
Specifically, for small sampling rates $r$ and small privacy loss parameters $\varepsilon$, applying certain simple sampling designs (simple random sampling with and without replacement, and Poisson sampling) before running an $\varepsilon$-DP mechanism reduces the privacy loss to approximately $r\varepsilon$.

However, most surveys conducted by statistical agencies use complex multistage sampling designs, potentially with different sampling strategies at the different stages. These designs are primarily used to increase the accuracy of the survey outputs or to reduce the survey's operational costs. For example, the Current Population Survey (CPS), one of the flagship surveys of the U.S. Census Bureau, uses a two-stage sampling design in which  stratified cluster sampling with probability proportional to size (PPS) is used to select clusters at the first stage and systematic sampling is used to sample households within clusters at the second stage \citep{bls-design}. There is no reason to believe that amplification effects for these complex designs are comparable to those obtainable for the simple designs discussed above. \citet{bunControllingPrivacyLoss2022} study the amplification effects for complex designs and find that amplification is small for most of the sampling designs used in practice. Their findings can be summarized as follows:
\begin{itemize}
    \item Cluster sampling using simple random sampling without replacement to draw the clusters offers negligible amplification in practice except for small $\varepsilon$ and very small cluster sizes.
    \item With minor adjustments, stratified sampling using proportional allocation can provide privacy amplification. For small $\varepsilon$, the amplification is still linear in the sampling rate up to a constant factor.
    \item Data dependent allocation functions such as Neyman allocation for stratified sampling will likely result in privacy degradation. (The effects will depend on the sensitivity of the allocation function.)
    \item With PPS sampling at the individual level, the privacy amplification will linearly depend on the maximum probability of inclusion (for small $\varepsilon$).
    \item Systematic sampling will only offer amplification if the ordering of the population is truly random. In all other cases, systematic sampling will suffer from the same effects as cluster sampling, leading to no amplification (assuming the ordering is known to the attacker).
\end{itemize}

 In practice this implies that for many multistage sampling designs, which typically start with (multiple stages of) stratified cluster sampling, amplification effects can generally only be expected from those stages at which individual units or households are selected (typically the last stage of selection).

\section{DP for Survey Weighted Estimates}\label{sectionSurveyWeights} 

As discussed in the introduction, the amount of noise that needs to be added to achieve a specific privacy loss $\varepsilon$ directly depends on the sensitivity of the statistic of interest. Intuitively, this makes sense. %
If the statistic changes substantially when one record is changed in the data it will be easier to infer that record's value from observing the statistic and thus more noise will need to be added to sufficiently protect that record. From a utility perspective, this implies that more reliable (less noisy) DP outputs can be expected from statistics with low sensitivity. Thus, a common strategy with DP is to identify estimation strategies with low sensitivity and replace very sensitive estimates with less sensitive alternatives, for example by using robust statistics \citep{dwork2009differential_lei,avella2021}.

When analyzing survey data, it is generally important to take the sampling design into account since the probabilities of selection typically vary between the units included in the sample. Unweighted estimates, especially those for descriptive statistics such as means and totals, will be biased whenever there are varying selection probabilities. To obtain unbiased estimates, each observation needs to be weighted by the inverse of its probability of selection. Hence, statistical agencies typically provide survey weights to enable researchers to take the survey design into account. In practice, these survey weights will also account for nonresponse and other data deficiencies such as undercoverage. (We will address this extra layer of complexity in the next section.) 

Using survey weighted estimates raises the question: how (if at all) does the sensitivity of a statistic change when the survey design is taken into account? To illustrate the possible impacts, let us assume the analyst is interested in estimating the mean of some variable $Y$ in the population using the sampled values $y_i$, $i={1,\ldots,n}$, where $n$ denotes the sample size. If the probabilities of selection were equal for all units, the sample mean would be an unbiased estimate for the population mean and its sensitivity would be $R/n$, where $R=\max(y_i)-\min(y_i)$ is the range of all possible values for $y_i$.\footnote{Throughout this section, we consider the bounded $\varepsilon$-DP setting. Similar arguments (with slightly different values for the sensitivity of a statistic) would apply for other settings.}
When dealing with unequal probabilities of selection, a popular estimator for the population mean is the Horvitz-Thompson estimator \citep{horvitz1952}: $\hat{\mu}_{Y}^{HT}=\sum w_i y_i/N$, where $w_i$ is the weight of unit $i$, for $i={1,\ldots,n}$ and $N$ is the size of the population. %
Note that we assume for simplicity that $N$ is known and does not need to be protected and $w_i$ is the design weight, i.e., it only accounts for the sampling design.

If we can treat the weights as fixed, the sensitivity of $\hat{\mu}_{Y}^{HT}$ is $\max(w_i)R/N$. Whether the maximum is over all units in the frame, over all units in the population, or over all possible counterfactual units, depends on which stages of the survey pipeline are treated as invariant as discussed in Section~\ref{DPInPipeline}. Note that for equal-probability designs all $w_i=N/n$ and thus the sensitivity of the Horvitz-Thompson estimator is the same as for the unweighted estimator. If $\max(w_i)>N/n$, the Horvitz-Thompson estimator will have larger sensitivity than the unweighted estimator. %

However, these discussions assume that the weights can be treated as fixed, that is, they do not change if a record changes in the database. For most sampling designs used in practice, such an assumption is unrealistic. For example, with sampling proportional to size (PPS), the $i$th record's probability of inclusion is given by $\pi_i=(n\cdot x_i)/N\cdot\bar{x}$, where $x_i$ %
is the value for unit $i$ of the measure-of-size variable $X$ that is used to improve the efficiency of the sampling design, and $\bar{x}=\sum_N x_i/N$ is the population mean of $X$. Changing the value of $X$ for a single record will change the probabilities of inclusion and thus the survey weights for all other records in the sampling frame. Therefore, the sensitivity will be larger compared to the setting with fixed weights as we no longer only need to consider the maximum possible change in a single record's value for $Y$. %
We also need to consider the impact of the weight change for all the other records even if their values for $Y$ don't change. %

A recently-proposed strategy to mitigate this potentially-substantial increase in sensitivity is to regularize the weights, as explored by \citet{seemanDifferentiallyPrivatePopulation2024}. (An extreme version of this strategy would set all weights to be equal; this could be justifiable if the increase in the privacy noise due to the weights dwarfs the bias introduced by ignoring the sampling design.) Another possible strategy is to treat the frame as invariant as discussed in Figure~\ref{figExamplePipeline3}. Frame invariance assumes any two neighboring datasets must always originate from the same frame and so can only differ at the sample level (or later). Thus, the probabilities of inclusion will be constant between neighboring datasets. However, treating the frame as invariant has two additional implications that need to be considered. First, fixing the frame implies that privacy amplification from sampling is no longer possible (we would need to have neighboring datasets at the frame level in order to achieve amplification). However, given the results of \citet{bunControllingPrivacyLoss2022}, this amplification is likely small in practice and thus the positive effects of reducing the sensitivity will tend to outweigh the negative effects of losing the amplification effect. On the other hand, fixing the frame will restrict the possible counterfactual input datasets to those that are consistent with the realized frame. Because this restriction will fix the survey weights, it might introduce strong constraints on the possible neighboring datasets, depending on the sampling design. As a consequence, the actual privacy guarantees for a frame invariant setting could be significantly weaker than the guarantees under a non-frame-invariant setting even for the same privacy loss parameter. How problematic this reduction in privacy is in real settings is currently an open question for research.

\section{DP and Weighting Adjustments}\label{sectionWeightAdjustments}

In practice, two adjustment steps are commonly applied to the design weights to correct for unit nonresponse and other data deficiencies such as over- or undercoverage in the sampling frame: nonresponse adjustments and calibration. Nonresponse is typically taken into account by modeling each survey unit's probability to respond and then multiplying the design weights with the inverse of the estimated response propensities. Calibration techniques rely on benchmarks known from other sources such as census data or large scale surveys such as the American Community Survey \citep{u.s.censusbureauAmericanCommunitySurvey2022}. These techniques can be used to adjust the survey weights in such a way that the survey weighted estimates will match the known benchmarks exactly. How these adjustment steps interfere with differential privacy has not been studied so far. (We are currently at an early stage of trying to address this problem.) However, both steps are data dependent, that is, they use information from the survey units for the adjustments. This implies that these steps cannot be ignored from a privacy perspective as the adjusted weights leak some personal information. Looking at the impacts on the sensitivity of the final statistic of interest (which uses the adjusted weights), similar problems as those discussed in the previous section will arise: changing one record in the database can potentially change the weight-adjustment factors for all other units in the survey. Thus, it seems imperative not to only account for these adjustment steps at the analysis stage. Better results in terms of the privacy--accuracy trade-off might be achieved if the weight-adjustment steps would be carried out in a differentially private way. More research is needed to better understand this trade-off. For example, it seems beneficial to identify robust adjustment strategies as less noise would be required to satisfy DP for these strategies.

In the particular case of post-stratification (which is a simple type of calibration), one such robust adjustment strategy has been proposed by \citet{cliftonPreliminaryReportDifferentially2023}. Another strategy would be to regularize the nonresponse and calibration weight adjustments. (This would be similar to the survey weight regularization strategy of \cite{seemanDifferentiallyPrivatePopulation2024}, discussed in the previous section.)

\section{DP and Imputation}\label{sectionImputation}

All survey data are plagued by item nonresponse as survey respondents are often unwilling or unable to respond to all survey questions especially if they request sensitive information. A common strategy to deal with this problem is to impute the missing values before analyzing the data. Imputation is especially helpful if the response process is selective, that is if it is not missing completely at random as defined by \citet{rubin1976inference}. In this case, using only the fully observed cases for the analysis would give biased results. However, imputations are always data dependent as they typically build a model based on the observed data and use this model to impute the missing values. As a consequence, the implications of imputation on the DP guarantees need to be considered regardless of whether or not the imputation procedure is included inside the data release mechanism. 
Some preliminary results for this problem are discussed in \citet{dasImputationDifferentialPrivacy2022}. 

Similar to the problem of weighting adjustments, there are two possible strategies to account for imputation under DP. The first strategy only considers the effects when analyzing the imputed data. The second strategy modifies the imputation routines to ensure that the imputations already satisfy DP. As \citet{dasImputationDifferentialPrivacy2022} have shown, the first strategy implies that in the worst case the sensitivity increases linearly with the number of imputed observations. This substantial increase of the sensitivity arises because changing one record in the database can potentially impact all of the imputed values. Whether the worst case applies depends on the analysis of interest and on the selected imputation procedure. Still, for statistical agencies offering pre-imputed datasets for accredited researchers, this strategy is not an option since they cannot anticipate which analyses might be performed on the imputed data.

The second strategy can break the dependence of the sensitivity on the number of imputed records---at least for certain imputation strategies. The key requirement for breaking this dependence is that the imputation model $m$ can be written as $D_{imp}^{(i)}\sim m(D_{obs}^{(i)},\hat{\theta})$, where $D_{imp}^{(i)}$ and $D_{obs}^{(i)}$ contain the imputed and observed variables for record $i$ and $\hat{\theta}$ denotes the model parameters estimated on the complete data. The model implies that, given $\hat{\theta}$, the imputed values of record $i$ only depend on the observed values of that record and not on any other record. If these requirements are met and the parameters $\theta$ of the imputation model are estimated using any suitable differentially private mechanism with privacy loss parameter $\varepsilon_1$, then, given any $\varepsilon_2$-differentially private mechanism used for analyzing the data, the overall privacy loss is given by $\varepsilon_1+\varepsilon_2$.
 
We note that the conditional independence assumption of the imputation model holds for many imputation methods, for example, parametric imputation models based on linear regression. However, it does not hold for hot-deck imputation, an imputation method commonly applied at statistical agencies.
  
\section{Discussion}\label{sectionConclusion}

DP is theoretically intuitive and elegant. It provides quantifiable and composable guarantees of privacy protection (although these guarantees have been subject to some confusion and misinterpretation, see \cite{tschantz2020sok}). By putting data privacy on a mathematical basis, it has supplied a calculus for reasoning about the protection offered by sophisticated data release mechanisms. %

Yet implementing DP mechanisms in practice often entails unforeseen complexities. %
In this chapter we have focused on some of the complexities that arise in the context of survey data. %
Many of the same complexities can also emerge in settings with data preprocessing steps or with multistage data collection (such as national censuses). The goals of this chapter are to draw attention to these complexities, review the current progress on addressing them, and spur renewed research activity to resolve those that remain outstanding.

We should emphasize at this point that we have explicitly focused on the complexities arising when using classical survey estimators, such as the Horvitz-Thompson estimator. These estimators all make use of the survey weights. However, there is an ongoing debate in economics about whether to include weights in econometric models \citep{solonWhatAreWe2015, gelmanStrugglesSurveyWeighting2007, mageeUseSamplingWeights1998, faiellaUseSurveyWeights2010}. Many researchers tend to ignore survey weights and instead account for the sampling design and nonresponse by adding control variables to their models. Most of the complexities discussed in this chapter (excluding the discussions on the increased sensitivity due to weighting in Sections~\ref{sectionSurveyWeights} and~\ref{sectionWeightAdjustments}) are still relevant in this context. Some of these complexities are further exacerbated when using complex econometric models, because computing the sensitivity for such models tends to be more difficult (as compared to simple statistics such as a Horvitz-Thompson estimator of a univariate mean). If we also want to account for earlier steps of the survey pipeline (e.g., nonresponse imputation or setting the sample size), calculating these models' sensitivity can become a major obstacle in practice. 

We should also emphasize that in this chapter we have only investigated the problem of releasing a single survey statistic, whereas in reality survey publications typically contain many outputs. For example, the American Community Survey asks respondents more than seventy questions, generates a dataset with over 64,000 variables, and produces more than 11 billion statistics annually \citep{uscensusbureauAmericanCommunitySurvey2025, jacobsenAmericasEssentialEconomic2023, uscensusbureauCensusAPIDatasets2022}. This introduces an additional complexity in adopting DP for survey data. When a record contributes to multiple statistics, the overall privacy loss $\varepsilon$ is typically equal to the sum of these statistics' individual privacy losses. This can quickly lead to either 1) very high levels of noise being added to all of the statistics; or 2) DP providing very weak---essentially meaningless---guarantees of protection. For example, given an overall privacy loss of $\varepsilon = 10$ (an upper limit for what can still be considered meaningful, see \cite{nearProgrammingDifferentialPrivacy}), releasing 64,000 (unweighted) count statistics would require infusing each of these counts with noise that has variance 81 million. %
An alternative would be to add noise that has variance 100,000, but this would result in an overall $\varepsilon \approx 286$---a value that is basically vacuous. (Setting the variance to be smaller would result in even larger values of the privacy loss parameter $\varepsilon$). While there are methods that achieve better trade-offs between $\varepsilon$ and the noise's variance \citep[e.g.,][]{li2015matrix, mckennaOptimizingErrorHighdimensional2018, xiaoOptimalScalableMatrix2023, abowd2022topdown}, they cannot currently handle a data product of the scale and complexity of the American Community Survey, and adjusting these methods to fit with the complications of survey data is a nontrivial, unresolved problem.

Having identified a multiplicity of challenges in obtaining DP---some of which may be unduly constraining---we suggest that future research investigates pragmatic modifications to ``completely-by-the-books'' implementations of DP. The goals of such modifications should be: to provide a solution that is feasible to implement; to retain the essence of DP even while not strictly satisfying DP; and to not unduly sacrifice the accuracy of the released data, nor the privacy of the data subjects, nor the resources of the statistical agency (in implementing the solution). %
Of course, any such modifications should be principled, in the sense that the associated risks to privacy are properly quantified and are outweighed by gains in data utility or implementability. Assessing the privacy risks of these modifications will likely involve a combination of theoretical and empirical analyses, and require measures of data privacy that lie outside the framework of DP. 

An example of one possible modification is the non-DP publication of a data-dependent sampling design. A description of a survey's sampling design is crucial information for data users. Yet, as outlined in Section~\ref{DPInPipeline}, if the sampling design was chosen with reference to the frame (as is often the case), then DP requires noise to be added to it before it can be published. Moreover, designing a DP mechanism to publish a sampling design will likely be difficult. On the other hand, it defies intuition that a simple description of a survey's sampling design should be disclosive of private information. This suggests it may be reasonable to modify the DP data release mechanism, allowing the sampling design to be released exactly (i.e., without noise) even while the other outputs are protected in line with the exact requirements of DP. But to justify this pragmatic violation of DP, the statistical agency should first address the questions: Can the risks associated with publishing a sampling design be quantified (without resorting to DP%
)? And when is it principled (in the sense given in the previous paragraph) to publish a sampling design as is, without privacy protection?

\bookmarksetup{startatroot}
\phantomsection\subsection*{Acknowledgments} 
JB gratefully acknowledges partial financial support from the Australian-American Fulbright Commission and the Kinghorn Foundation. JD gratefully acknowledges support from Cooperative Agreement CB20ADR0160001 with the U.S. Census Bureau. The views expressed in this chapter are those of the authors and not those of the Census Bureau or any other sponsor.

\begingroup
\sloppy
\printbibliography
\endgroup

\end{document}